\documentclass{jltp}
\usepackage{graphicx} % uncomment this line to include the graphicx package
\def\be{\begin{equation}}
\def\ee{\end{equation}}

\title{ Monte Carlo Simulations of Spin-Diffusion in a 2-D Heisenberg Paramagnet }
\author{ Robert Ragan, Kurt Grunwald, and Chris Glenz }
\address{ Department of Physics\\
University of Wisconsin-La Crosse \\
La Crosse, WI 54601 USA}
\runninghead{R. Ragan {\it et al.}}{Monte Carlo Simulations of Spin-Diffusion  ...}
\begin{document}
\maketitle
\begin{abstract}
We study spin diffusion and spin waves in paramagnetic quantum crystals 
(solid $^{3}$He, for example) by direct simulation of a square lattice 
of atoms interacting via a nearest-neighbor Heisenberg exchange 
Hamiltonian.  Recently, Cowan and Mullin have used a moments method to study spin transport at 
arbitrary polarizations.  We test their 
analytic results by calculating the statistical spin correlation function 
from molecular dynamics simulations using a Monte Carlo algorithm to average over initial spin configurations.  Since it is not practical to diagonalize the $S={1\over 2}$ exchange Hamiltonian for a lattice which is of sufficient size to study long-wavelength (hydrodynamic) fluctuations, we instead study the $S \rightarrow \infty$ limit and treat each spin as a vector with a classical equation of motion.  We compare our simulations with the assumptions of the moments method regarding the short-time behavior and the long exponential tail of the correlation function.  We also present our numerical results for the polarization dependence of the longitudinal spin diffusion coefficient and the complex transverse spin diffusion coefficient.
\end{abstract}
\section{INTRODUCTION}
The Leggett-Rice (LR) equation\cite{Legg70} was originally derived for paramagnetic Fermi fluids, but it has been shown that it applies to the spin dynamics of any polarized system interacting via quantum exchange.\cite{Meyer90} In the linear regime, the LR equation describes conventional diffusion of longitudinal Fourier modes $S^{z} \propto e^{i {\bf q}\cdot {\bf r}-D_{\parallel}q^{2}t}$, where $D_{\parallel}$ is the longitudinal diffusion coefficient, as well as damped transverse spin waves $S^{+} \equiv S^{x}+ iS^{y} \propto e^{i {\bf q}\cdot {\bf r}-D^{+}q^{2}t}$, characterized by a complex diffusion coefficient $D^{+}$. The transverse diffusion coefficient can be written in terms of a bare diffusion coefficient as

\be
D^{+}=\frac{D_{\perp}}{1-i\mu M}
\label{Dperp}
\ee
where $D_{\perp}$ is real and the denominator takes into account the precession of the spin current about the molecular field with quality factor $\mu M = {\rm Im}D^{+}/{\rm Re}D^{+}$.

Cowan and Mullin {\it et al.}\cite{Cowan89}$^{-}$\cite{Cowanun} (CM) have applied this description to the spin dynamics of paramagnetic $^{3}$He in its hcp phase, with a nearest-neighbor pair exchange Hamiltonian, and have used a moments method, originally due to Redfield\cite{Red} and deGennes,\cite{deG} to calculate the polarization dependence of the transport coefficients. At zero polarization $P=0$ they found ordinary spin diffusion ($\mu M=0$, $D_{\perp}=D_{\parallel}\sim \omega a^{2} $), where $\omega$ is the exchange frequency and $a$ is the lattice spacing. For $P \rightarrow 1$ they found that all the transport coefficients diverge, but with $D_{\perp}/\mu M$ finite, as one would expect for undamped spin waves. Over a wide range of $P$ they found $D_{\perp}\approx D_{\parallel}$, although the two diverge differently as $P\rightarrow 1$. 

In order to test this picture we carried out a numerical study of spin diffusion on a 2d square lattice with the Hamiltonian 
\be
H=-\sum_{i} BS^{z}_i-\sum_{i,j}J{\bf S}_{i} \cdot {\bf S}_{j}
\label{Ham}
\ee
were the second sum is over nearest neighbors. We shall restrict our attention to the weakly paramagnetic case, with $J/T \rightarrow 0$ but $B/T$ remaining finite. Since a lattice of $N$ spins involves $2^{N}$ quantum states, it is not practical to diagonalize the $S={1\over 2}$ exchange Hamiltonian for a lattice which is of sufficient size to study long-wavelength (hydrodynamic) fluctuations. We instead study the $S \rightarrow \infty$ limit and treat each spin as a vector ${\bf S}({\bf r})$ with a classical equation of motion, \cite{Wang}$^{,}$\cite{Dzh} 
\be
\frac {d}{dt}{\bf S}({\bf r})=\omega \sum_{{\bf r}^{\prime}} {\bf S}({\bf r}) \times {\bf S}({\bf r}^{\prime})
\label{motion}
\ee
where the sum is over nearest neighbors. For convenience, we have moved to the Larmor frame, and we set the exchange frequency $\omega=J/\hbar$ and the vector magnitude of the spins $|{\bf S}|$ equal to $1$. 
\begin{figure}
\centerline{\includegraphics[height=2.5in]{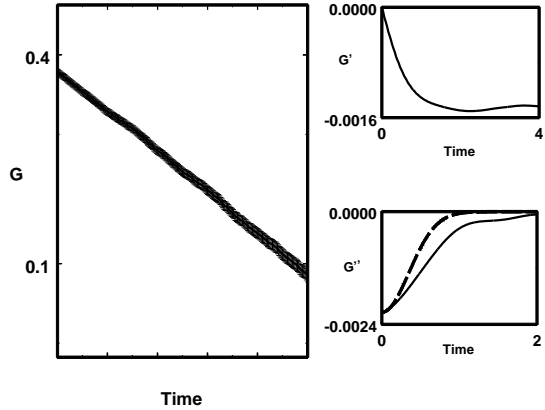}}
%
%\framebox[5in]{\rule[1.125in]{0in}{1.125in}}
%\makebox[5in]{\rule[1.125in]{0in}{1.125in}}
\caption{Log plot of the longitudinal spin correlation function $G^{z}_{q}(t)$ at zero polarization. The slope is equal to $-D_{\parallel}q^{2}$ from which a value of $D_\parallel= 0.51 \pm 0.01$ (in units of $\omega a^{2}$)is obtained. To the right, the short time behaviors of the first and second derivatives are also shown. For comparison, the dashed line in the $G^{\prime\prime}$ plot is the Gaussian ansatz used by CM.\cite{Cowan89}}
\label{fig:Figure1}
\end{figure}
The longitudinal diffusion coefficient was calculated by simulating the motion of a hydrodynamic ($q \rightarrow 0$) fluctuation $S^{z}_{q}(t)$. We define the correlation function\cite{Forster}
\be 
G^{z}_{q}(t)=\langle S^{z}_{q}(t) S^{z*}_{q}(0)\rangle 
\label{corr}
\ee
which is assumed to have an exponential decay $\propto e^{-D_{\parallel}q^{2}t}$ as $t\rightarrow \infty$. The brackets indicate an average over initial conditions sampled from an equilibrium ensemble. 
\section{NUMERICAL SIMULATIONS}
We shall outline the procedure for calculating $D_{\parallel}$. First, an initial spin configuration was obtained by generating a lattice of randomly oriented spins ${\bf S}_{i}$ with a Boltzman distribution $e^{(B/T)S^{z}_{i}}$. The exchange energy was neglected in the Boltzman factor, since in a weakly paramagnetic system the spins are basically uncorrelated, although their motion is not. The spin configuration was then allowed to evolve according to Eq.\ref{motion}, and the spin correlation function $G^{z}_{q}(t)$ was calculated from the definition Eq.\ref{corr}. The average was then performed over $N$ initial configurations were $N=1000$ was typical. The size of the lattice was $L_{x}\times L_{y}= 64\times 8$ spins with periodic boundary conditions. The length $L_{x}$ of the lattice was chosen so that the wave number $q=2\pi/L_{x} = 0.098$ was small. To insure that the lattice was a 2-$d$ system, $L_{y}$ was chosen so that the mean free time $\tau \equiv D_{\parallel}/(\omega a)^{2}$ was shorter than the time $L_{y}^{2}/(D_{\parallel}\pi^{2})$ to diffuse across the width of the lattice. The equation of motion was integrated with a second-order modified Euler method, with time step $\Delta t = 0.01$. The typical motion of a spin in a time step was half a degree or so. The conserved quantities such as the total energy, the polarization, and the individual spin magnitudes were found to be constant to a high degree of accuracy. 

In Fig.1 the correlation function $G^{z}_{q}(t)$ is shown for $P=0$. A similar procedure was used to calculate $D_{\perp}$ and the quality factor $\mu M$ from $S^{+}_{q}$. Some typical results are shown in Fig. 2 for $P=0.673$.

%\begin{figure}
%
%\centerline{\includegraphics[height=2.5in]{Figure2.EPS}}
%
%\framebox[5in]{\rule[1.125in]{0in}{1.125in}}
%\makebox[5in]{\rule[1.125in]{0in}{1.125in}}
%\caption{Log plot of magnitude of the transverse spin correlation
% $|G^{+}_{q}(t)|$ for $P=0.673$. The inset shows the oscillations of its real 
%and imaginary parts.}  
%\label{fig:Figure2}
%\end{figure}
\begin{figure}
\centerline{\includegraphics[height=2.5in]{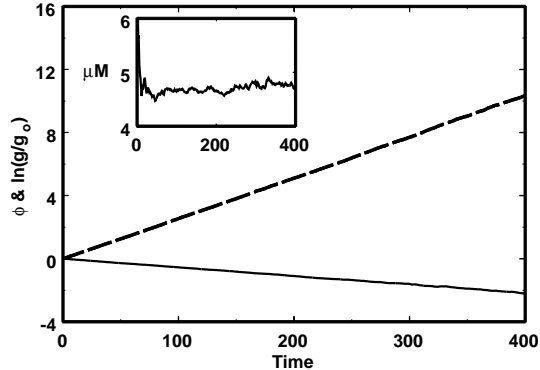}}
%
%\framebox[5in]{\rule[1.125in]{0in}{1.125in}}
%\makebox[5in]{\rule[1.125in]{0in}{1.125in}}
\caption{Plot of $\log |G^{+}_{q}(t)|$ (solid line)and the phase (dashed line) from the previous figure. A quality factor of $\mu M= 4.6\pm 0.1$ is found from the ratio of the slopes of the two lines. The inset shows the instantaneous value of $\mu M$. }
\label{fig:Figure3}
\end{figure}
\begin{figure}
\centerline{\includegraphics[height=2.5in]{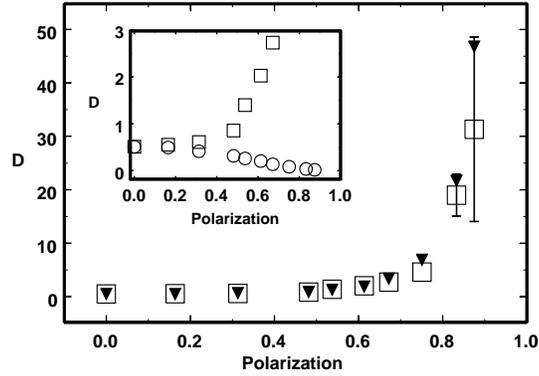}}
%
%\framebox[5in]{\rule[1.125in]{0in}{1.125in}}
%\makebox[5in]{\rule[1.125in]{0in}{1.125in}}
\caption{Polarization dependence of $D_{\parallel}$ (squares) and $D_{\perp}$ 
(triangles). The error bars are smaller than the size of the symbols, except for the two $D_{\parallel}$ points at $P>0.8$. The inset shows $D_{\parallel}$ and the effective transverse 
diffusion coefficient ${\rm Re}D^{+}$ for comparison with Fig.1 of Ref.\onlinecite{King96}.}
\label{fig:Figure4}
\end{figure}
\begin{figure}
\centerline{\includegraphics[height=2.5in]{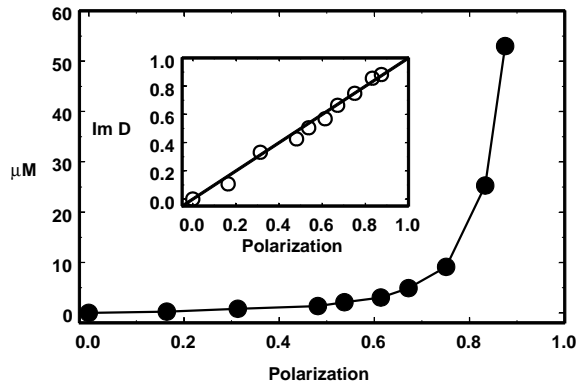}}
%
%\framebox[5in]{\rule[1.125in]{0in}{1.125in}}
%\makebox[5in]{\rule[1.125in]{0in}{1.125in}}
\caption{Polarization dependence of the quality factor. The inset shows ${\rm Im}D^{+}$. Obviously, ${\rm Im}D^{+}\approx P \omega a^{2}$. Since the spin wave frequency is ${\rm Im}D^{+}q^{2}$, this supports the notion of a molecular precessional field. It also agrees with the mean field results of CM, and agrees with the exact solution at $P=1$.}
\label{fig:Figure5}
\end{figure}
\section{RESULTS AND DISCUSSION}
The polarization dependence of the transport coefficients are shown in Fig.3-4.  The results are very similar to results of CM obtained with a moments method. As expected, all the transport coefficients diverge as $P \rightarrow 1$, although $D_{\perp}/\mu M$ remains finite. There was no indication of anisotropy in the diffusion coefficients for $P<0.8$. At higher polarizations the divergence of $D_{\parallel}$ required prohibitively large lattices and calculation times. This was not a problem in the transverse case, since ${\rm Re} D^{+} \rightarrow 0$ as $P \rightarrow 1$.  
Some disagreement with CM is expected since they considered a $3d$ spin-$1/2$ system using a mean field approximation. When our transport coefficients are plotted vs. $1-P^{2}$ on a log-log plot the resulting curve fits are linear and indicate a divergence $\propto (1-P^{2})^{3.0\pm 0.3}$. This disagrees with CM who found an exponent of $-1$ for the transverse coefficients and $-1/2$ for $D_{\parallel}$.\cite{King96}   

However, there is remarkable agreement at low polarization. For example, at zero polarization CM obtained $D_{\parallel}=D_{\perp}=0.79$ for a $2d$ square lattice.\cite{Cowan89}  This is higher than our value of $0.51$, but when comparing classical and quantum results one should make the substitution $S^{2}\rightarrow S(S+1)$ with $S=1/2$. We omit the details of the analysis here, but this increases our value by a factor of $\sqrt{3}$ to $0.87$. Also, for $3d$ hcp they obtain $\mu M/P =2.3$ 
as $P\rightarrow 0$,\cite{King96} which is higher than our value of 1.5. The quantum correction above cancels for this quantity, but from connectivity arguments, one would expect the molecular field in a $3d$ system to be stronger by a factor $3/2$, which would increase our value to $2.3$, in perfect agreement with CM. 

Finally, our simulations with classical spin vectors support some of the assumptions of the moments method. For long times, hydrodynamic fluctuations were found to decay exponentially. For short times, the second derivative of the correlation function was well-described by a Gaussian ansatz and decayed on a time scale $\tau \equiv D_{\parallel}/(\omega a)^{2}$ (See Fig.1).

\section*{ACKNOWLEDGMENTS}
The authors thank W. Mullin and B. Cowan for useful discussions. This research was supported by NSF grant DMR-0071706, an NPACI allocation, and an AEG grant from Sun Microsystems, Inc.

\end{document}